\documentclass[a4paper]{jpconf}
\usepackage{graphicx}

\def\be{\begin{equation}}
\def\ee{\end{equation}}

\def\ergs{{\rm\,erg\,s^{-1}}}

\def\ergs{\rm \,erg\,s^{-1}}
\catcode`\@=11 
\def\@versim#1#2{\vcenter{\offinterlineskip
        \ialign{$\m@th#1\hfil##\hfil$\crcr#2\crcr\sim\crcr } }}
\def\lsim{\mathrel{\mathpalette\@versim<}}
\def\gsim{\mathrel{\mathpalette\@versim>}}
\def\mpy{M_\odot \ {\rm yr^{-1}}}
\begin{document}

\title{Accretion models of Sgr A*}

\author{Feng Yuan$^{1,2}$}

\address{$^1$ Shanghai Astronomical Observatory, Chinese Academy of
Sciences, 80 Nandan Road, Shanghai 200030, China}
\address{$^2$ Joint Institute for Galaxy and Cosmology (JOINGC) of SHAO
and USTC}

\ead{fyuan@shao.ac.cn}

\begin{abstract}

The supermassive black hole in the center of our Galaxy, Sgr A*, is unique
because the angular size of the black hole is the largest in the sky
thus providing detailed boundary conditions on, and much less freedom for,
accretion flow models. In this paper I review advection-dominated accretion 
flow (ADAF; another name is radiatively inefficient accretion flow) models for 
Sgr A*. This includes the developments and
dynamics of ADAFs, and how to explain observational results
including the multi-waveband spectrum, radio polarization, IR and X-ray 
flares, and the size measurements at radio wavebands. 
\end{abstract}

\section{Introduction}

The center of our Galaxy provides the best evidence to date for a
massive black hole (e.g., Sch\"odel et al. 2002; Ghez et al. 2003),
associated with the compact radio source, Sgr A* (see, e.g., Melia \&
Falcke 2001). Since the original discovery of Sgr A* in 1974,
there have been intensive efforts in both observational and theoretical 
aspects, with dramatic progresses in the past few years. The reason why we
are so interested in this object is because of its proximity, which
allows us to determine observationally the dynamics of gas quite 
close to the BH, providing unique constraints on theoretical 
models of accretion flows. 

Before introducing the model, I first briefly review the main 
observational results of Sgr A*. 
As shown by the data points in Fig. 2, its radio spectrum consists of two
components.  The component below 86 GHz has a spectrum $F_{\nu} \propto 
\nu^{0.2}$,
while the high frequency component, the ``submm bump'', has a spectrum
$F_{\nu} \propto \nu^{0.8}$ up to $\sim 10^3$ GHz (e.g., 
Falcke et al. 1998; Zhao et al. 2003). Variability at centimeter and millimeter
wavelength are detected with a timescale ranging from hours to years 
with amplitude of less than 100\% (Zhao et al. 2003, 2004; 
Herrnstein et al. 2004; Miyazaki et al. 2004). High level of 
variable linear polarization fraction ($\sim 2\%-10\%$)
at frequencies higher than $\sim 150$ GHz puts a
rotation measure upper limit of $7\times 10^5 {\rm rad~m^{-2}}$, which argues
for a low density at the innermost region of ADAF (e.g., Aitken et
al. 2000; Bower et al. 2003; Marrone et al. 2006; Macquart et al. 2006;
Quataert \& Gruzinov 2000). 

At IR wavelength, the source is highly variable. Genzel et al. (2003)
detected Sgr A* at 1.6-3.8 $\mu$m, with a factor
of $\sim 1-5$ variability on timescales of $\sim 10-100$ min.
Similarly, at 3.8 $\mu$m, Ghez et al. (2004) found that the flux
changes by a factor of 4 over a week, and a factor of 2 in
just 40 min. If describing the IR spectrum with a power-law, Gillessen et al. 
(2006) found that the spectral index is correlated with the instantaneous flux
but Hornstein et al. (2006) found it remains constant during the flare
process. 

Sgr A* has been convincingly detected in the X-rays
(Baganoff et al. 2001, 2003; Goldwurm et al. 2003). 
The X-ray emission has two distinct components.  In
``quiescence,'' the emission is soft and relatively steady, with a
large fraction of the X-ray flux coming from an extended region with a
diameter $\approx 1.4''$ (Baganoff et al. 2001, 2003).  Several times
a day, however, Sgr A* has X-ray ``flares'' in which the X-ray luminosity
increases by a factor of a few -- 50 for roughly an hour. 
For the most flares, the spectrum is
hard, with a photon index of $\Gamma=1.3^{+0.5}_{-0.6}$.
{\em XMM}, however, detected a very bright and soft flare
with $\Gamma=2.5^{+0.3}_{-0.3}$ (Porquet et
al. 2003). Recent several multiwavelength campaigns found that there 
is no time lag between the IR and X-ray flares (Eckart et al. 2004, 2005; 
Yusef-Zadeh et al. 2006). This strongly
suggests a common physical origin. The short timescale 
argues that the emission arises quite close to the BH,
within $\sim 10 R_S$ (where $R_S$ is the Schwarzschild radius).
Sgr A* is extremely dim overall, with a bolometric luminosity 
of only $L\approx 10^{36}\ergs
\approx 3\times 10^{-9}L_{\rm Edd}$.

\section{Accretion flow models: from ADAF to RIAF}

Consider a steady axisymmetric accretion flow. The dynamics of the flow are
described by the following height-integrated differential equations,
which express the conservation of mass, radial momentum, angular momentum,
and the energy of electrons and ions (e.g., Narayan, Mahadevan \& Quataert 1998):

\be \dot{M}=-4\pi R H \rho v = \dot{M}_{\rm out}\left(\frac{R}{R_{\rm
out}} \right)^s, \ee
\be v\frac{dv}{dr}=-\Omega_{\rm k}^2 r+\Omega^2 r-\frac{1}{\rho}\frac
{dp}{dr}, \ee 
\be v(\Omega r^2-j)=\alpha r \frac{p}{\rho}, \ee
\be \rho v
\left(\frac{d \varepsilon_e}{dr}- {p_e \over \rho^2} \frac{d
\rho}{dr}\right) =\delta q^++q_{ie}-q^-, \ee
\be \rho v
\left(\frac{d \varepsilon_i}{dr}- {p_i \over \rho^2} \frac{d \rho}{dr}
\right) =(1-\delta)q^+-q_{ie}=-(1-\delta)\alpha p r
\frac{d\Omega}{dr}-q_{ie}. \ee
The quantities have their usual meaning. Specifically 
$s$ describe the strength of the outflow and $\delta$ the fraction of the turbulent
energy which directly heats electrons. The radiative processes include
synchrotron and bremsstrahlung emission and their Comptonization. After the
dynamics of the accretion flow is solved, we can calculate the emitted spectrum
and polarization fraction and compare with the observed ones. 

The above set of equations hold for any accretion models including, e.g., 
the standard thin disk, except that the standard thin disk is one-temperature due to
the strong coupling between electrons and  ions so the two energy equations are
unified into one. The ADAF solution is one set of solution in the regime 
of $\dot{M}\lsim \alpha^2\dot{M}_{\rm Edd}$ where $\dot{M}_{\rm Edd}\equiv
10L_{\rm Edd}/c^2$ is the Eddington accretion rate\footnote{There are 
four solutions to the eqs. (1)-(5), two ``cold'' ones and two ``hot''
ones. The former includes the standard thin disk and 
slim disk; the latter includes ADAF and LHAF. See Fig. 6 in 
Yuan 2001 and Fig. 1 in Yuan 2003 for details.}. An ADAF has many
distinct properties compared to the standard thin disk. The gas temperature
is much higher, the flow is optically thin and geometrically thick.
Perhaps most importantly, the radiative efficiency of an ADAF is very low,
$\eta_{\rm ADAF} \sim 0.1 \dot{M}/(\alpha^2\dot{M}_{\rm Edd})$ (for details, see
Narayan \& Yi 1994, 1995; Narayan, Mahadevan \& Quataert 1998). This is the
reason why ADAF models are so successful in explaining the dim feature of Sgr A*
(Narayan et al. 1995, 1998; Manmoto et al. 1997). 

There has been a significant change in the theoretical understanding 
of ADAFs over the past few years.  First, global, time-dependent, numerical
simulations reveal that very little mass available at large radii actually
accrets onto the black hole and most of it is lost to a magnetically
driven outflow or circulates in convective motions (e.g., Stone,
Pringle, \& Begelman 1999; Hawley \& Balbus 2002), i.e, $s>0$ in eq. (1). 
The physical reason for the
presence of outflow is that since very little energy 
is radiated away, the Bernoulli parameter of an ADAF 
is almost zero\footnote{The self-similar solution of
Narayan \& Yi (1994) gives a positive Bernoulli parameter of the flow. However,
Nakamura (1998) argue that the Bernoulli parameter 
in an ADAF should be mostly negative. This is confirmed by my global solution 
of ADAFs.}, much larger than that in the standard thin disk, therefore 
the gas is earlier to escape to infinity once they 
are perturbed (Narayan \& Yi 1994; Blandford \& Begelman 1999). 

In the early version of an ADAF, $\delta$
{\em is assumed} to be very small, $\delta\approx
 10^{-2}-10^{-3}$, i.e., the turbulent
energy only heats ions. However, there has been large 
theoretical uncertainties in how electrons are heated in the accretion flow
(Quataert \& Gruzinov 1999) and a larger $\delta$ seems to be 
physically more plausible since processes like magnetic reconnection
is likely to heat electrons directly. A large $\delta$ is 
also required from the 
observational side. Quataert \& Narayan (1999) show that 
when outflows are present, to produce the correct amount of emission 
of Sgr A*, the electrons in the accretion flow should be hotter 
than in the early version 
of ADAF to compensate for the decrease of radiation due to the decrease
of density. Stated another way, $\delta$ must be larger, 
$\delta \lsim 1$. Sometimes,
the ``updated'' version of ADAF with $s>0$ and $\delta \lsim 1$
is called radiatively inefficient accretion flow (RIAF). In this review we
simply use the original term: ``ADAF''.

When $\delta\approx 10^{-2}$, almost all of the turbulent 
energy $q^+$ is stored in the flow as the entropy of ions and very little 
is transferred into electrons through Coulomb
collision $q_{ie}$, i.e., $q_{ie}\ll q^+$.
We generally think this is why the radiative
efficiency of ADAF is so low.
If $\delta \lsim 1$, which means that a large amount of turbulent energy will
heat electrons directly, what is the radiative efficiency of an ADAF?
In this case, as shown by Fig. 1, 
almost all of the turbulent energy heating electrons
is stored in electrons as entropy, i.e., the electrons are advection-dominated.
This ensures the low efficiency of an ADAF even though $\delta$ is large. 
We would like to emphasize that with the increase of accretion rate, the 
flow is no longer advection-dominated and the 
radiative efficiency rapidly increases (ref. Fig. 7 in 
Narayan, Mahadevan \& Quataert 1998 and Fig. 6 in Yuan 2001). 

\begin{figure}[h]
\vspace{1pc}
\includegraphics[width=16pc]{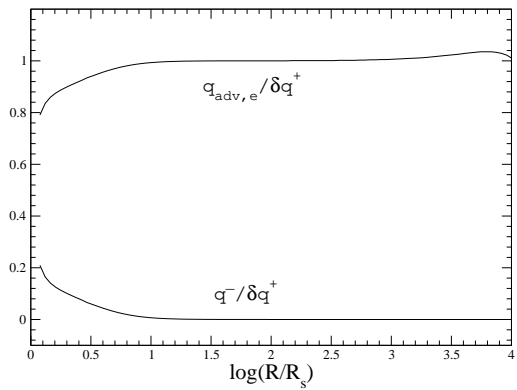}
\hspace{2pc}
\begin{minipage}[b]{16pc}
\caption{\label{label2}The energy relationship of electrons in 
the ADAF in Sgr A* (with $\delta \approx 0.55$).}
\end{minipage}
\end{figure}

\section{Application to Sgr A*}

\subsection{Outer boundary conditions}

The outer boundary conditions include
the temperature, density, mass accretion rate,
 and angular momentum at the outer boundary. The outer boundary is 
determined by the Bondi radius. For an uniformly
distributed matter with an ambient density $\rho_0$ and sound speed $c_s$, the
Bondi radius of a black hole of mass $M$ is $R_{\rm Bondi}
\approx GM/c_s^2$ and the accretion rate is $\dot{M}_{\rm Bondi}\approx
4\pi R_{\rm Bondi}^2\rho_0c_s$.
{\em Chandra} observations of the Galactic center detect extended diffuse emission
within $1-10^{"}$. The inferred gas density and temperature are 
$\approx 100~ {\rm cm}^{-3}$ and $\approx 2 $ keV on $1^{"}$ scales 
(Baganoff et al. 2003). The corresponding Bondi radius $R_{\rm Bondi} \approx 
0.04 {\rm pc}\approx 1^{"}\approx 10^5R_s$ and the Bondi accretion rate is 
$\dot{M}_{\rm Bondi} \approx 10^{-5} \mpy$. 
Given that the accreted gas should have some angular momentum (see below),
the accretion rate should be somewhat smaller. The recent 3D numerical
simulation for the accretion of stellar winds on to Sgr A* by Cuadra et al. (2006) 
obtains $\dot{M}\approx 3\times 10^{-6}\mpy$. 
If gas were accreted at this rate onto the black hole via the standard thin 
disk, the expected luminosity would be $L\approx 0.1\dot{M}_{\rm Bondi}
c^2\approx 10^{41}\ergs$, 5 orders of magnitude higher than the observed 
luminosity. This is the strongest argument against a thin disk in Sgr A*.

The numerical simulation indicates that the angular momentum is  
large, with the circularization radius of being about $10^4R_s$ 
(Cuadra et al. 2006). This result casts 
doubt on the ``small angular momentum'' assumption of spherical accretion 
models (Melia 1992; Melia et al. 2001).

\subsection{Explaining the quiescent state}

\begin{figure}[h]
\includegraphics[width=17pc,angle=0]{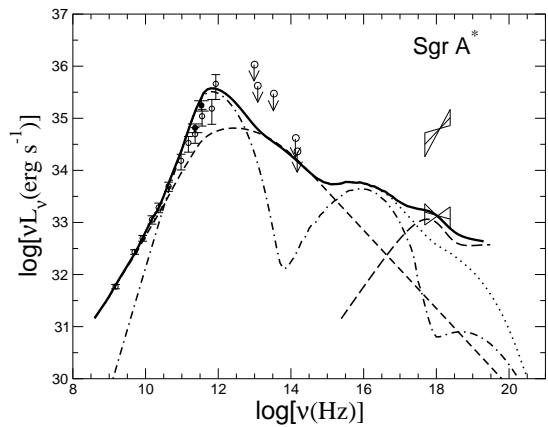}
\hspace{4pc}
\begin{minipage}[b]{15pc}
\caption{\label{label} ADAF model for the quiescent state emission from
Sgr A*. The dot-dashed line is the
synchrotron and SSC emission by thermal electrons; the dashed line is
the synchrotron emission by non-thermal electrons.
The dotted line is the total synchrotron and SSC emission while the
solid line includes the bremsstrahlung emission from the outer parts
of the ADAF (shown by the long-dashed line). Taken from YQN03.}
\end{minipage}
\end{figure}

A number of authors have applied ADAF models to explain the observations of
Sgr A*. In the early applications (e.g., Narayan et al. 1995, 1998; 
Mahadevan et al. 1997), they neglect outflows and the direct heating of electrons
by turbulence, i.e., $s=0$ and $\delta=10^{-2}$ are adopted. These work can 
roughly account for the observed low luminosity and spectrum of Sgr A*. However,
such models can't explain the observed high linear polarization.
This is because the predicted density is too high and the rotation measure is
$\sim 10^{10}{\rm radm^{-2}}$ in the region where 
high-frequency radio emission is produced, much higher than the upper limit 
of $7\times 10^{5}{\rm rad~m^{-2}}$ obtained by Marrone et al. (2006). 
This problem can be solved if 
we take into account the presence of outflows, i.e, $s>0$. 
In this case, most of the accretion gas is lost to the outflow thus 
the density at the innermost region is much lower. 

Another problem with the original ADAF model is that it under-predicts the
low-frequency radio emission significantly. This can be solved by introducing 
another component to the model. One possibility is a jet (Falcke \& Markoff
2000; Yuan, Markoff \& Falcke 2002). Alternatively, some
fraction of electrons in the ADAF may be in nonthermal distribution, due to 
acceleration processes in the accretion flow such as weak 
shocks and turbulent dissipation.
Their synchrotron emission can explain the radio ``excess''.
Fig. 2 shows the
modeling result of an ADAF model with outflow and nonthermal electrons, 
taken from Yuan, Quataert \& Narayan (2003, hereafter YQN03). 
The parameters are $\dot{M}_{\rm out}\approx 10^{-6}
\mpy, s=0.27, \delta=0.55$. The nonthermal electrons are assumed to be in a 
power-law distribution with the spectral index of $p=3.5$. The injected energy
in nonthermal electrons is equal to a fraction $1.5\%$ of the energy in thermal 
electrons. The accretion rate close to the black hole is 
$\dot{M}\approx 4\times 10^{-8}\mpy \ll \dot{M}_{\rm out}$. As shown in the 
figure, the submm bump is mainly due to the synchrotron emission from thermal
electrons, the low-frequency radio and IR emissions
 from the synchrotron emission from nonthermal electrons, while the 
extended X-ray emission from the bremsstrahlung emission from the region around
the Bondi radius.

\begin{figure}[h]
\includegraphics[width=34pc]{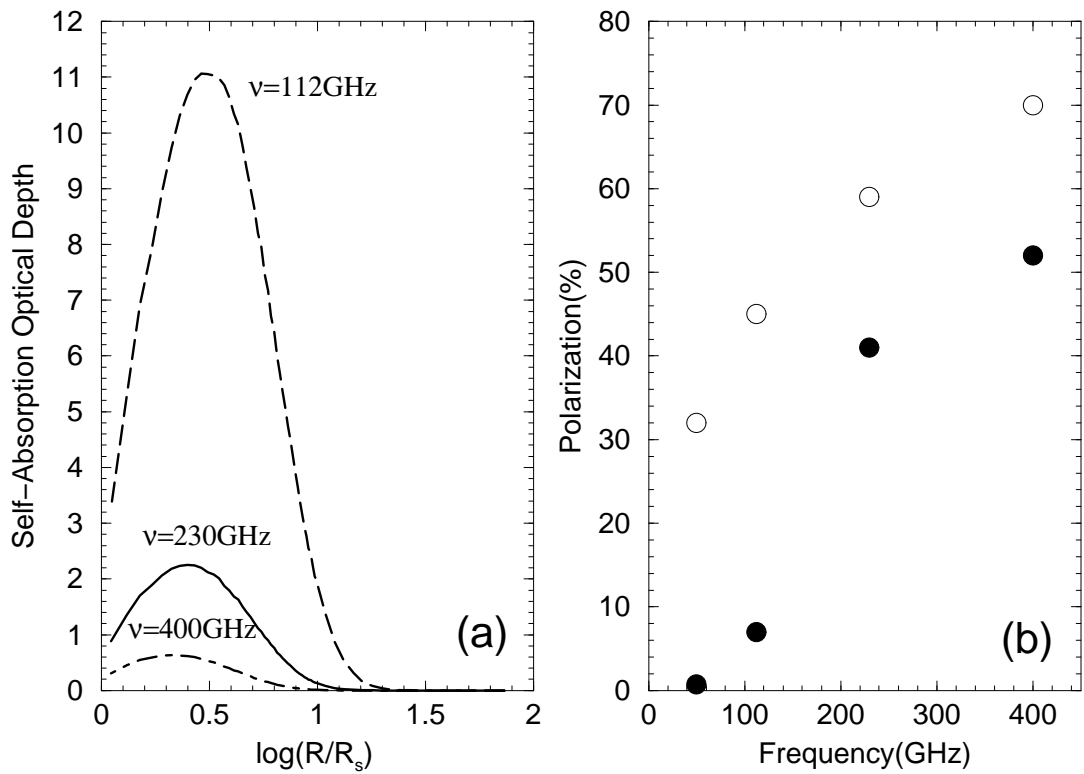}
\caption{\label{label} (a) Synchrotron self-absorption optical depth of
thermal electrons as a function of radius; (b) Polarization fraction from thermal
electrons with (filled circles) and without (open circles) Faraday rotation.
 These values are upper limits because a uniform magnetic field is assumed. 
Taken from YQN03.}
\end{figure}

Since most of the accretion gas is lost into the outflow, the rotation measure
is very small, ${\rm RM}\approx 10^7$ and $\lsim 5\times 10^5{\rm rad~m^{-2}}$ 
at 230 GHz if we integrate along the equatorial plane and rotation axis, 
respectively. Note that they should be regarded as upper limits since we assume
the magnetic field is fully coherent and point along the line of sight while 
the magnetic field should be predominantly toroidal. 

As noted above, the submm emission is due to thermal
electrons. Since the linear polarization of optically thick thermal
synchrotron emission from a uniform medium is suppressed by
$\exp[-\tau]$, where $\tau$ is the synchrotron optical depth, one also
needs to check whether the model can roughly account for the magnitude
of the observed linear polarization ($\lsim 10\%$). YQN03 
have calculated the linear polarization produced
by the thermal electrons in their models, including optical depth
effects and Faraday rotation. Fig. 3a  shows $\tau$ as a function 
of radius for three frequencies and Fig. 3b shows by open circles the degree of
linear polarization as a function of frequency when Faraday rotation
is neglected. From the figure we see that thermal electrons can 
readily account for the observed level of linear polarization from Sgr A*.

\subsection{Understanding the flares}

Markoff et al. (2001) showed that
the flares are probably due to enhanced electron heating or
acceleration, rather than a change in the accretion rate onto the black hole.
The IR and X-ray flares in Sgr A* may be analogous with 
solar flares, in which magnetic energy is converted into thermal
energy, accelerated particles, and bulk kinetic energy, due to magnetic 
reconnection. This speculation is confirmed by the 3D MHD numerical simulation
by Machida et al. (2003). 
Unfortunately, even for solar flares, some important aspects such as the 
detailed acceleration mechanism and the energy distribution of
heated/accelerated electrons remains unclear (Miller 1998).

\begin{figure}[h]
\vspace{1pc}
\begin{minipage}{18pc}
\includegraphics[width=18pc]{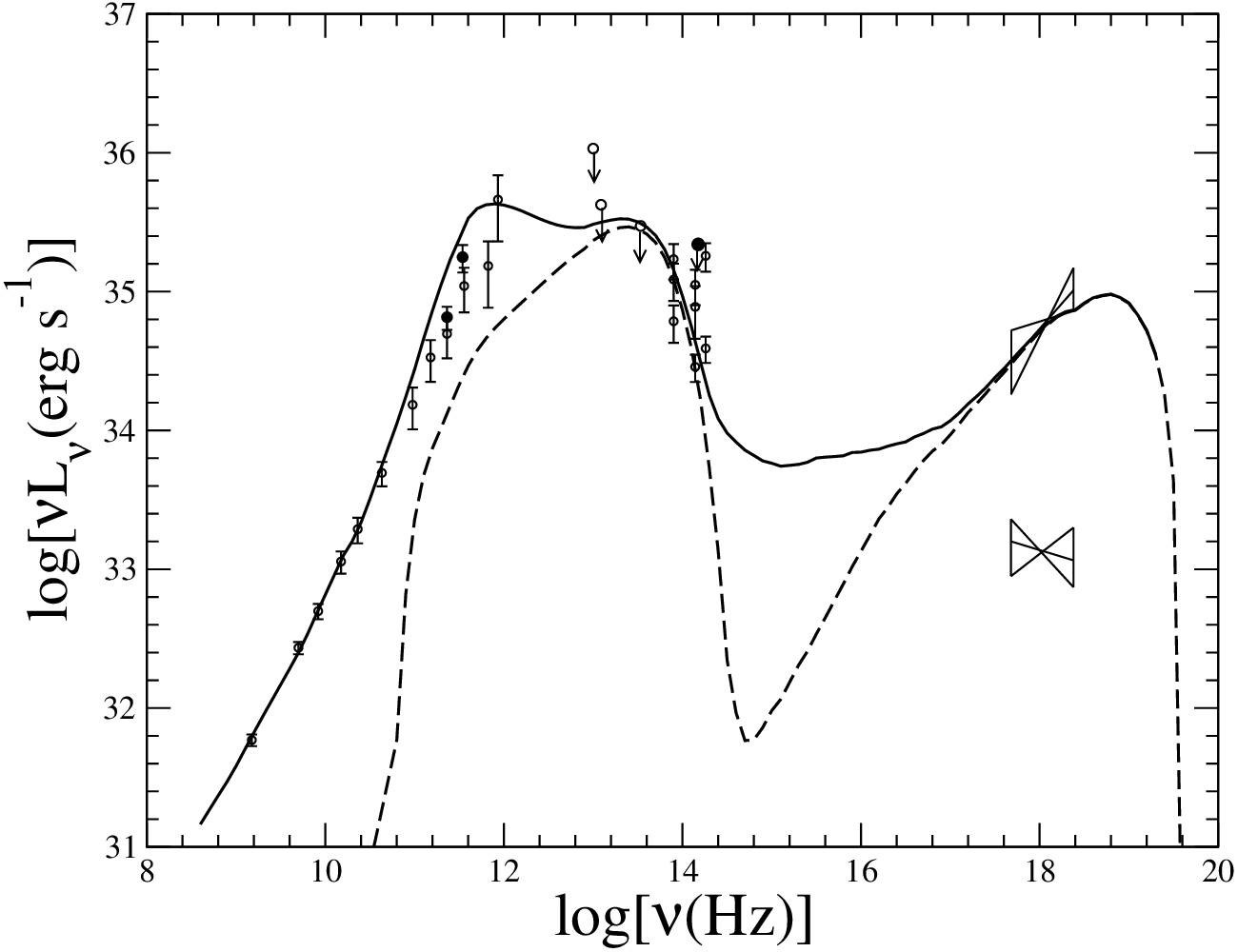}
\caption{\label{label} Synchrotron+SSC model for the IR and X-ray flares
in Sgr A*. Taken from YQN03.}
\end{minipage}
\hspace{2pc}
\begin{minipage}{18pc}
\includegraphics[width=18pc]{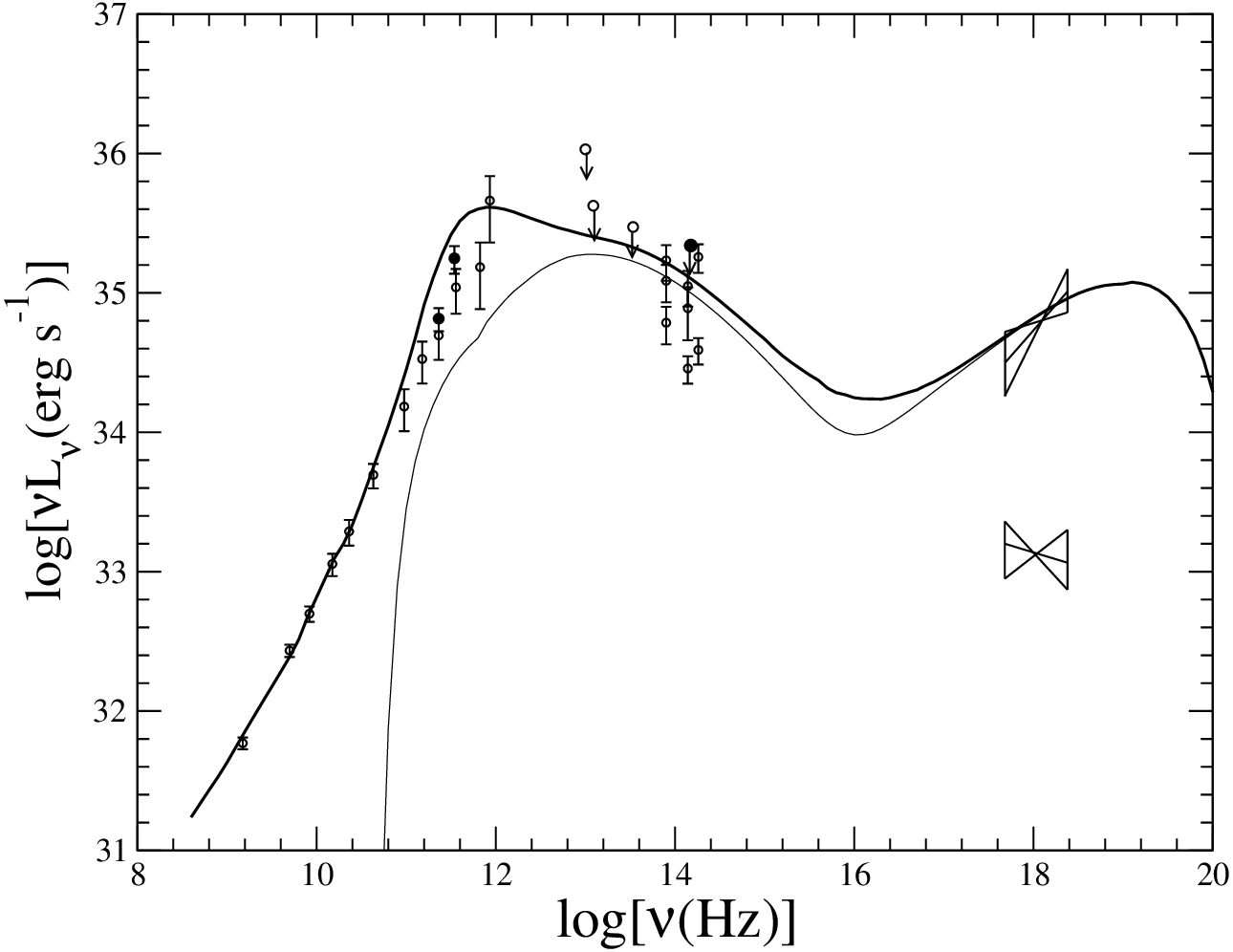}
\caption{\label{label1} Synchrotron model for the IR and X-ray flares
in Sgr A*. Taken from Yuan, Quataert \& Narayan 2004.}
\end{minipage}
\end{figure}

Based on this idea, YQN03 and Yuan, Quataert \& Narayan (2004) propose that
during the flare events, magnetic reconnection occurs in the innermost region
of the ADAF. As a result, some electrons are heated/accelerated into a
thermal/power-law distribution. Two models are proposed
to explain the IR and X-ray flares in Sgr A*. In the first ``synchrotron+SSC''
model shown by Fig. 4, the electrons are 
accelerated into a single power-law distribution
with the maximum Lorentz factor of $\gamma_{\rm max} \sim $ several hundred. 
The IR flare is due to the synchrotron emission 
while the X-ray flares due to the up-scattering of the IR photons 
by these electrons. In the second ``synchrotron''
model shown by Fig. 5, it is assumed that,
some electrons are {\em heated} into a higher
temperature while some electrons are {\em accelerated} into a (hard) power-law 
distribution with $\gamma_{\rm max}\sim 10^6$. The synchrotron emission of 
these two types of electrons are responsible for the IR and X-ray 
emissions, respectively\footnote{Liu et al. (2006)
propose that the electrons are in a {\em thermal} distribution after some stochastic
acceleration process. The synchrotron emission and its inverse Compton scattering
produce the observed IR and X-ray flares.}. 
I would like to emphasize that both models
can explain the observations well and there are no strong arguments against
any one given our poor knowledge of particle acceleration.

\subsection{Using the size measurement to test the models}

Recent radio observations by the VLBA at 7 and 3.5 mm produce the
high-resolution images of Sgr A*, and detect its
wavelength-dependent sizes (Bower et al. 2004;
Shen et al. 2005). The measured size
of images of Sgr A* at 7 mm is $0.712^{+0.004}_{-0.003}$ mas by Bower et al. (2004),
and 0.724 $\pm 0.001$ mas and $0.21^{+0.02}_{-0.01}$ mas at 7 and
3.5 mm, respectively by Shen et al. (2005). This provides an additional test to 
accretion models of Sgr A*. 

\begin{figure}[h]
\includegraphics[width=36pc]{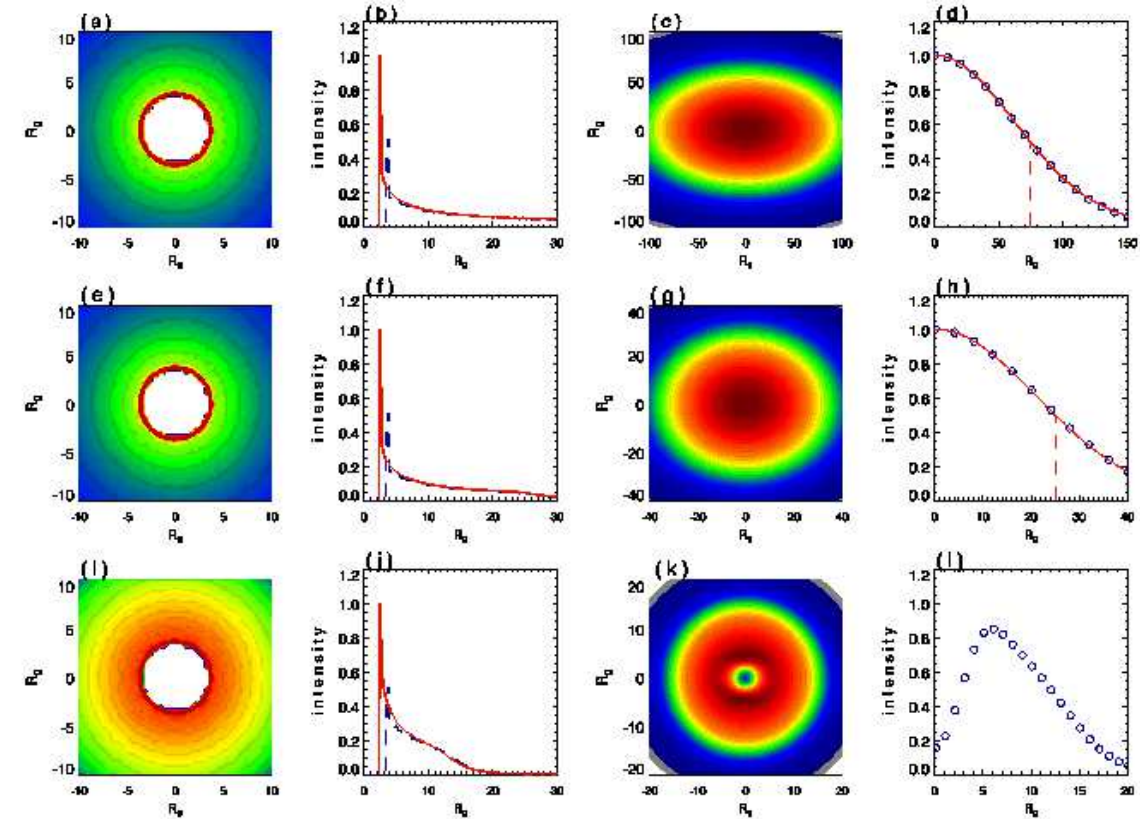}
\caption{\label{label} The images and sizes of Sgr A* at 7, 3.5,
and 1.3 mm from top to bottom. The
first column shows the ``input'' intensity distribution. The 
second column shows the intensity profile calculated from
the YQN03 model, with (dashed) and without (solid) GR effect taken into account. 
The third column shows the simulated image after the
interstellar scattering is taken into account. 
The fourth column shows the intensity profile of the simulated
image (dots) and the Gaussian fit (line).
The vertical dashed line shows the location of the FWHM. At 1.3
mm, the simulated profile can't be fit by a Gaussian, and thus no
FWHM is indicated. Taken from Yuan, Shen \& Huang 2006.}
\end{figure}

Yuan, Shen, \& Huang (2006) calculate the predicted size of the 
YQN03 model. The most important point they realize is that
since the intrinsic intensity profile of the ADAF is not a Gaussian, as shown
by the dashed lines in the second column of Fig. 6, thus it is impossible to 
directly compare the predicted intrinsic size with observation. 
Rather, they calculate the predicted intrinsic
intensity profile of the ADAF models, then taking into account the 
scattering broadening toward the Galactic center to obtain the {\em simulated} 
image. They then compare the simulated image with the observed one. The third
column in Fig. 6 shows the simulated image at three wavelengths and
the fourth column shows the fitting results of the intensity
profile by a Gaussian. The obtained sizes of the simulated images 
are $0.729^{+0.01}_{-0.009}$ mas and $0.248^{+0.001}_{-0.002}$
mas at 7 mm and 3.5 mm, respectively. The simulated size at 7 mm is in good agreement
with the observed value by Shen et al. (2005) within the error
bars but slightly larger than the observed size by Bower et al.
(2004); the size at 3.5 mm is a little larger than the observation
of Shen et al. (2005). Given that the size of the source may be
variable (Bower et al. 2004) and the uncertainties in the
calculations, they conclude that the
predictions of the YQN03 model are in reasonable agreement with
the observations. They predict that GR effects may be detectable at 1.3 mm.

There are two alternative models of Sgr A*, 
namely the jet model of
Falcke \& Markoff (2000) and the coupled jet-ADAF model of Yuan,
Markoff \& Falcke (2002). One difference between these two
models is that in the former the
radio emission above $\sim 86 $GHz is produced by the nozzle of
the jet while in the latter the contribution of the ADAF is
significant. In the jet-ADAF model, the contribution of the emission from 
the ADAF can dominate over that from the jet under suitable parameters. 
Moreover, if the ``old'' ADAF in this model is ``updated'' (i.e, including outflow), 
the only difference between the jet-ADAF model and the 
YQN03 model is the origin of the radio
emission below $\sim$ 86 GHz. In this case, the predicted size of Sgr A*
by the jet-ADAF model will be consistent with the
observations. On the other hand, the predicted sizes at 3.5 mm 
by the jet model is $\gsim 0.04$ mas (Falcke \& Markoff 2000),
much smaller than the observed value, so some modifications
of the model are required (Markoff et al. in preparation). 

\section{Summary}
 
The supermassive black hole in our Galactic center represents a unique opportunity
to probe the physics of accretion, especially at extremely low accretion rates.
In this review, I first briefly introduce the dynamics and evolution of the 
ADAF. I then have tried to argue that this
model can provide a reasonable explanation to most of the current observations.

{\em Chandra} observations tell us the density and temperature of the 
hot gas at $\sim 1^{"}\sim 0.04$ pc. This radius happens to be the Bondi radius
where the gas is captured by the gravity of the center black hole and starts to be
accreted. The Bondi accretion rate, $\dot{M}_{\rm Bondi}\sim 10^{-5}\mpy$
provides a good estimation to the real accretion rate. As a comparison, the
numerical simulation gives $\dot{M}\sim 3\times 10^{-6}\mpy$. 
Since the bolometric luminosity of Sgr A* is only $10^{36}\ergs$,
the radiative efficiency should be very low, $\sim 5\times 10^{-6}$. 
The standard thin disk model (Shakura \& Sunyaev 1976)
is therefore ruled out immediately.

``Old'' ADAF models can naturally explain such a low efficiency (e.g., Narayan
et al. 1995). However, these models fail to explain the high
linear polarization at submm waveband because the density and further, the
rotation measure, are too large. On the other hand, theoretical studies of
ADAFs also indicates the presence of outflow (e.g., Stone, Pringle 
\& Begelman 1999; Blandford \& Begelman 1999). This feature is taken into 
account in the ``new'' ADAF models (or RIAF; YQN03). The YQN03 model
can explain the quiescent state spectrum and the polarization, as shown by 
Figs. (2)\&(3). The accretion 
rate close to the horizon is only $4\times 10^{-8}\mpy$. So the low radiative 
efficiency of the ADAF in Sgr A* is partly because of the outflow, which
contributes a factor of $\sim 10^{-2}$, and partly because of the energy
advection, which contribute a factor of $\sim 5\times 10^{-4}$. The IR and 
X-ray flares are explained by the synchrotron  and inverse-Compton
emissions from the heated/accelerated
electrons during the magnetic reconnection events in the innermost region of the 
ADAF (Figs. 4\&5). Finally, the YQN03 model satisfactorily pasts the 
test of recent observations of the size of Sgr A* at 3.5 and 7 mm wavebands (Fig. 6).
Moreover, this model predicts that the observation at 1.3 mm should be able to 
detect GR effects (Fig. 6). 

\ack
This work was supported in part by the
One-Hundred-Talent Program and the National Natural Science
Foundation of China (grants 10543003)

\end{document}